\documentstyle [12pt,a4p,epsfig,amsmath,multicol]{article}
\newcommand {\rbf}  {\overline{r}_f}
\newcommand {\vbf}  {\overline{v}_f}
\newcommand {\abf}  {\overline{a}_f}
\newcommand {\sbf}  {\overline{s}_f}

\newcommand {\rbl}  {\overline{r}_l}
\newcommand {\vbl}  {\overline{v}_l}
\newcommand {\abl}  {\overline{a}_l}
\newcommand {\sbl}  {\overline{s}_l}

\newcommand {\vbc}  {\overline{v}_c}
\newcommand {\abc}  {\overline{a}_c}
\newcommand {\sbc}  {\overline{s}_c}

\newcommand {\vbb}  {\overline{v}_b}
\newcommand {\abb}  {\overline{a}_b}
\newcommand {\sbb}  {\overline{s}_b}
\newcommand {\sbQ}  {\overline{s}_Q}
\newcommand {\gbl}  {\overline{g}_b^L}
\newcommand {\gbr}  {\overline{g}_b^R}

\newcommand {\Afbl}  {A_{FB}^{0,l}}
\newcommand {\Afbc}  {A_{FB}^{0,c}}
\newcommand {\Afbb}  {A_{FB}^{0,b}}
\newcommand {\tpol}  {$\tau$-polarisation}

\newcommand {\sleff}  {\sin^2\Theta_{eff}^{lept}}
\begin{document}
\hspace*{8cm} {UGVA-DPNC 1998/09-179 September 1998}
\newline
\hspace*{10.5cm} hep-ph/9809292
\vspace*{0.6cm}
\begin{center}
{\bf  Z-DECAYS TO b QUARKS AND THE HIGGS BOSON MASS}
\end{center}
\vspace*{0.6cm}
\centerline{\footnotesize J.H.Field}
\baselineskip=13pt
\centerline{\footnotesize\it D\'{e}partement de Physique Nucl\'{e}aire et 
 Corpusculaire, Universit\'{e} de Gen\`{e}ve}
\baselineskip=12pt
\centerline{\footnotesize\it 24, quai Ernest-Ansermet CH-1211 Gen\`{e}ve 4. }
\centerline{\footnotesize E-mail: john.field@cern.ch}
\baselineskip=13pt
 
\vspace*{0.9cm}
\abstract{ A model independent analysis of the most recent averages of
 precision electroweak data from LEP and SLD finds a 3$\sigma$ 
 deviation of the parameter $A_b$ from the Standard Model prediction.
 The fitted value of $m_H$ shows a strong dependence on the inclusion
 or exclusion of b quark data, and the Standard Model fits have poor
 confidence levels of a few percent when the latter are included. 
 The good fits obtained to lepton data, c quark data and the
 directly measured top quark mass, give $m_t = 171.2_{-3.8}^{+3.7}$~GeV
 and indicate that the Higgs boson mass is most likely less than 200 GeV.
\vspace*{0.9cm}
\normalsize\baselineskip=15pt
\setcounter{footnote}{0}
\renewcommand{\thefootnote}{\alph{footnote}}
\newline 
{\it Keywords ;} Standard Electroweak Model, LEP and SLD data, Z-decays,
 b quark couplings, top quark and Higgs boson masses
\newline

\vspace*{0.4cm}

\newpage
  The dependence of the indirect upper limit on the Higgs boson mass ($m_H$) on
 the particular choice of high precision electroweak data has been discussed
 previously in the literature~\cite{x1,x2}, as well as the problem of combining
 direct and indirect limits on $m_H$~\cite{x3}. It has also recently been noticed
that a model independent analysis of Z decay data yields effective b-quark
 couplings that differ by three standard deviations or more from the
 predictions of the Standard Electroweak Model (SM)~\cite{x4,x5}. In this paper
 a model independent analysis similar to that described in Refs.[4,5] is performed
 on the latest averages of precision electroweak data~\cite{x6}. Assuming only
lepton universality and the validity
of perturbative QED and QCD corrections, several parameters,
 directly sensitive to electroweak quantum
corrections, and so to the mass of the top quark ($m_t$) as well as $m_H$, are
extracted from the data. The SM predictions, including the recently calculated
$O(g^4m_t^2/M_W^2)$ two-loop corrections~\cite{x7} as implemented in the program
package ZFITTER5.10~\cite{x8} are then fitted to the data to determine indirectly
the values of $m_t$ and $m_H$, as well as the overall confidence level (C.L.) for
 agreement of the fit with the SM. 
 \par The six parameters extracted are: $A_l$, $A_c$, $A_b$, $\sbl$, $\sbc$ and
 $\sbb$ where the subscripts $l$, $c$, and $b$ denote charged leptons (assuming
 universality), c quarks and b quarks respectively. These parameters, equivalent
 to the effective couplings of leptons, c quarks and b quarks, contain all the
 high precision information on $m_t$ and $m_H$ from Z decay measurements. They
 are defined, in terms of the effective vector ($\vbf$) and axial vector ($\abf$)
 coupling constants of the fermion (charged lepton or quark) $f$, by the relations:
 \begin{equation}
    A_f \equiv \frac{2 (\sqrt{1-4 \mu_f}) \rbf}{1-4 \mu_f+(1+2 \mu_f) \rbf^2},
    \end{equation} 
 where 
\[  \rbf \equiv \vbf/\abf, \]
 and 
 \begin{equation}
 \overline{s}_f \equiv (\overline{a}_f)^2(1-6 \mu_f)+(\overline{v}_f)^2.
 \end{equation} 
 The parameter  $\mu_f = (\overline{m}_f(M_Z)/M_Z)^2$, where
$\overline{m}_f(Q)$ is the running fermion mass at the scale $Q$, can be set
to zero for $f=l,c$ to sufficient accuracy, while for b quarks 
$(\overline{m}_b(M_Z)/M_Z)^2 = 1.0 \times 10^{-3}$~\cite{x9}. The extraction
of $A_f$ from the various asymmetry and polarisation measurements is
 straightforward~\cite{x4,x5}. The quantities $\sbf$ are derived, from the
 measured partial widths, using the relations: 
 \begin{equation}
     \sbl = (\abl)^2+(\vbl)^2 = \frac{12 \pi \Gamma_l}{\sqrt{2} G_{\mu} M_Z^3}
     \frac{1}{(1+\frac{3 \alpha(M_Z)}{4 \pi})},
 \end{equation}  
and  
\begin{equation}
\sbQ = \sqrt{\frac{2 \pi}{3}}\frac{R_Q \Gamma_Z}{G_{\mu} M^2_Z}
\frac{\sqrt{R_l \sigma^0_h}}{C_{QED}^Q C_{QCD}^Q}~~~(Q=c,b),
\end{equation}
where 
\[ C_{QED}^Q =  1+ \frac{3 (e_Q)^2}{4 \pi} \alpha (M_Z),  \]
and 
 \[ C^Q_{QCD} = 1+ C_1^Q(\frac{\alpha_s(M_Z)}{\pi})+C_2^Q(\frac{\alpha_s(M_Z)}
{\pi})^2. \]
 The coefficients $C_1^Q$ and $C_2^Q$ may be found in Refs.[4,5] to which the
reader is referred for all questions of notation. The world average value 
$ \alpha_s(M_Z) = 0.118(5)$~\cite{x10,x11} is used in Eqn.(4).
The parameters $A_f$ (derived from asymmetry or polarisation measurements)
 and $\sbf$ (derived from the partial width of Z decay into $f \bar{f}$
 pairs) have the advantage of uncorrelated experimental errors,
 thus simplifying the error treatment of the fitting procedure.
\par 
  The averages of electroweak parameters used in the analysis are presented
in Tables 1 and 2\footnote{Unless otherwise stated, all errors quoted are the
quadratic sum of the experimental statistical and systematic errors}.
 Table 1 contains quantities directly sensitive to 
$m_t$ and $m_H$, while other essential parameters are found in Table 2.
\par Before performing any fits to the data, the overall consistency of 
 the different asymmetry and~\tpol~ measurements is checked by calculating,
in each case, the effective leptonic weak mixing angle\footnote{
 Note that $\sleff$, $\rbl$ and $A_l$ are strictly equivalent physical
 quantities related to each other by one-to-one mappings}:
\begin{equation}
 \sleff = \frac{1-\rbl}{4}.
\end{equation}
 The comparison of different measurements of $\sleff$ is shown in Fig.1. When
measurements of the same quantity by different experiments are in good agreement,
as is the case for $A_{FB}^{0,e}$, $A_{FB}^{0,\mu}$ and $A_{FB}^{0,\tau}$ the 
averages over experiments are shown. For the case of~\tpol, where the consistency
 of the different experiments is poor~\cite{x4,x5} the results of the 
individual LEP experiments are shown. For the purely leptonic measurements
(including $A_{LR}$), where only the assumption of lepton universality is needed
to extract $\sleff$, the weighted average value of 0.23125(23),
 (whose $\pm 1 \sigma$ region is indicated by the shaded band in Fig.1) is in 
 good agreement with the individual measurements ($\chi^2 = 8.1$, 7 $dof$, $\rm{C.L.}
 =0.32$). One may note, in particular, the good agreement of the $A_{LR}$ value
with the weighted average. Indeed, only the ~\tpol~ measurements of DELPHI and
OPAL are more than one standard deviation from the average value. Also shown
in Fig.1 is the value of $\sleff$ derived from $\Afbb$. In this case, besides
lepton universality, it is necessary to assume the SM values of the b quark
effective couplings. Including this datum in the average gives a somewhat
worse consistency $\chi^2$ ($\chi^2 = 13.3$, 8 $dof$, $\rm{C.L.}=0.10$) and changes 
the weighted average value by $1.1\sigma$ to 0.23151(20). The deviation of the
$\sleff$ value derived from $\Afbb$ from the leptonic average amounts to
$2.2\sigma$.  
\par The six parameters $A_f$, $\sbf$ ($f=l,c,b$) extracted by the model 
independent analysis are presented in Table 3, where they are compared with the SM 
prediction for $m_t=174$~GeV and $m_H=100$~GeV. All the parameters show good
agreement with the SM except for $A_b$ which lies 3.0$\sigma$ below the 
expectation. The effective vector and axial-vector coupling constants of leptons,
c quarks and b quarks are compared with the SM predictions in Table 4. Both
$\vbb$ and $\abb$ show deviations of about 3$\sigma$ from the SM; all other
couplings are in good agreement. The right-handed (R) and left-handed (L) 
effective couplings of the b quarks:~$\gbr = (\vbb-\abb)/2$, $\gbl = (\vbb+\abb)/2$
are found to have the values:
\[ \gbr = 0.1050(90),~~~~~~\gbl = -0.4159(24) \]
to be compared with the respective SM predictions of 0.0774 and -0.4208, As also
observed for the 1996 data averages~\cite{x4,x5} a somewhat larger deviation (3.1$\sigma$)
is seen for $\gbr$ than for $\gbl$ (2.0$\sigma$). In Fig.2 are shown the 68$\%$, 95$\%$ 
and 99$\%$ C.L. contours of the fit of
 $\gbr$ and $\gbl$ to $A_b$ and $\sbb$. The SM
prediction lies just outside the 99$\%$ contour.
\par Results of SM fits for $m_t$ and $m_H$ to different subsets of the data are presented
in Table 5 and Fig.3. In the fits the correlations between the errors on $A_l$ and $A_Q$,
 ($Q=c,b$)
 derived from the measurements of $A_{FB}^{0,Q}$, and those between $\sbb$ and $\sbc$ are
 taken into account. The direct SLD measurements of $A_c$, $A_b$ give separate 
uncorrelated\footnote{The correlation coefficients between $A_c$, $A_{FB}^{0.c}$
 and $A_b$, $A_{FB}^{0.b}$ are small (0.07,0.03 respectively~\cite{x15}) and are
 neglected in the fits}
 contributions to the $\chi^2$. The errors on $m_t$, $m_H$ quoted in Table 5
are at the 1$\sigma$ level on the individual parameters i.e. one parameter is 
fixed at the value corresponding to the minimum of the combined fit, the other being varied 
 till $\chi^2 = \chi_{min}^2+1$.
 The 95$\%$ C.L.  upper limits on $m_H$ are derived from a two-sided 90$\%$
 confidence interval given  by $\chi^2 <  \chi_{min}^2+2.7$.  
 The leptonic data
 gives a fitted value of $m_t$ about 2$\sigma$ below the directly measured (CDF and D0) value
 of $173.8 \pm 5.0$ GeV~\cite{x6}, and 
$m_H = 25.8$ GeV. Adding the c quark data (second row of Table 5) leaves the fitted values 
of $m_t$ and $m_H$ almost unchanged and a good overall fit (C.L.= 56$\%$)is
obtained. Combining 
instead the lepton and b quark data (the third row of Table 5) reduces the fitted value
of $m_t$ by only 3 GeV, but increases the fitted value of $m_H$ to 42 GeV. This fit has
a poor C.L. of only 1.3$\%$, reflecting the large deviation of $A_b$ from the SM
prediction. The results of the fit to all data from Table 3 are given in the last row of
Table 5. The fitted value of $m_t$ is almost the same as for the fit to the lepton and
b quark data only, $m_H$ increases to 45 GeV and the C.L. improves to 3.9$\%$. The 68$\%$
C.L. contours in the $m_H$ versus $m_t$ plane for the fits to the lepton and c quark data
( second row of Table 5, contour A)
 and all data ( fourth row of Table 5,  contour B)
 are shown in Fig.3. Also shown in Fig.3 is the current 
95$\%$ C.L. lower limit on the Higgs boson mass of 89.8 GeV from the direct search results of
the four LEP experiments~\cite{x12}, and the directly measured  value of $m_t$.
 Inclusion or exclusion of the b quark data has a strong
effect on the 1$\sigma$ contour. Indeed, the 1$\sigma$ contour A lies
almost entirely within
the region for $m_H$ forbidden at 95$\%$ C.L. by the direct search result. Also shown in 
Table 5 are the results of fits to $m_t$ only, fixing $m_H$ at 200 GeV.
 For the fits to the lepton and lepton and c quark
data the fitted values of $m_t$ increase, 
 whereas the C.L.s of the fits decrease, falling to only 2.5$\%$ for the
lepton data. A similar behaviour is seen in the fits to lepton and b quark data and all
data, however the already poor C.L.s of the fits with $m_t$ and $m_H$ free become even
worse: 0.3$\%$ and 2.0$\%$ respectively.
 \par All the fits with $m_t$ and $m_H$ free shown in Table 5 give values of $m_t$
about 10 GeV below the directly measured value. Also all the 95$\%$ C.L. upper limits 
on $m_H$ calculated as described above and shown in Table 5, are less than 90 GeV and
 so are only 
marginally consistent with the direct lower limit. However, the $m_H$ limits also depend
strongly on the value of $\alpha(M_Z)$. In order to take into account all available experimental information, the fits
to the lepton and c quark data and all data are repeated including also in the definition
of $\chi^2$ the directly measured value: $m_t=173.8 \pm 5.0$ GeV~\cite{x6}. The fits are also repeated with
$\pm1\sigma$ variations in the value of $\alpha(M_Z)$. The results of these fits are
presented in Table 6.  
\par A very stable value of $m_t$ of $m_t=171\pm1$ GeV, consistent with the directly 
measured value, is found for all six fits. The 95$\%$ C.L. upper limits on $m_H$ are
 found to be $105_{-46~-25}^{+70~+36}$ GeV for leptons and c quarks only, and
 $156_{-64~-33}^{+84~+44}$ GeV
 for all data. The quoted errors on the limits  correspond, respectively,
 to $\mp 1 \sigma$ variations of $\alpha(M_Z)$
 about the value quoted in Table 2, and $\pm 1 \sigma$ variations of
 $m_t$ about the fitted values shown in Table 6.  The `maximal'\footnote{i.e. the value 
obtained by adding linearly the +1$\sigma$ errors due to $\alpha(M_Z)^{-1}$ and $m_t$.}
 value of 284 GeV, in the case the
 b quark data is included, agrees well with the 95$\%$ C.L. upper limit of 280 GeV
 from the CERN Electroweak Working Group quoted in Ref.[6]. The corresponding
 maximal value for leptons and c quarks only is 211 GeV. It is clear from Table 6 that
 consistency with the direct lower limit on $m_H$ strongly disfavours values of 
$\alpha(M_Z)$ near to the +1$\sigma$ experimental limit. 
 The sensitivity of Higgs mass limits to inclusion or exclusion of the b quark
data is evident from the fits shown in Tables 5 and 6. Inclusion of the b quark data raises
the 95$\%$ C.L. from $\simeq 105$ GeV to $\simeq 156$ GeV for $\alpha(M_Z)^{-1} = 128.896$.
 However, the quality of the SM fits is then poor with C.L.s of only a few $\%$. 
 A careful examination of the `pulls'( datum - best fit value) of the fits shows
 that, although the shift in $m_H$ is entirely due to the deviation in $A_b$,
the bias is actually produced by variation of the parameter $A_l$ (very sensitive
 to $m_H$), rather than $A_b$ (completely insensitive to $m_H$ with the present
 experimental errors). The measured quantity $A_{FB}^{0,b} = 3 A_l A_b /4$ lies
 below the SM value corresponding to the purely leptonic data due to the
 low value of $A_b$. This deviation is reduced in the fit by increasing $m_H$
 so that $A_l$ is reduced. $A_b$ itself is essentially unchanged by the 
 variation in $m_H$. However, the leptonic data strongly constrains the 
 possible downward variation of $A_l$. This explains both the large `$2\sigma$'
 `pulls' observed for the 3 quantities $A_b$, $A_{FB}^{0,b}$ and $\sin^2 \theta_{eff}
^{lept}$ from $A_{LR}$ in the global fit shown in Ref.[6] and the poor
 confidence levels of all the fits including b quark data shown in Tables 5 and 6.  
\par The previous papers discussing the sensitivity of the indirect limit on $m_H$ to
 the data from different electroweak measurements focussed on the apparent
 incompatiblity of $\sleff$ derived from the SLD $A_{LR}$ measurement, both with the 
 other LEP measurements and the direct LEP lower limit on $m_H$. The analysis of 
 the authors of Ref.[2], based on the data available at the end of 1995~\cite{x14}
 also considered the effect of $R_b$, which, at that time, differed from the SM
 prediction by about 3$\sigma$. Gurtu~\cite{x1}, using the same 1996 data set~\cite{x15}
 as in the previous model independent analysis~\cite{x4,x5} of the present author, concluded
 that the $A_{LR}$ measurement must either have been subject to a very large
 statistical fluctuation, or an unknown systematic error which was then treated by the
 Particle Data Group rescaling procedure~\cite{x16}. Actually, a more careful 
 examination of the data~\cite{x4,x5} shows that the situation for the 1996 data 
 was quite similar to that of the current data
 \footnote{The values of $\sleff$ derived from $A_{LR}$ and $\Afbb$ in 1996 were 0.23060(48)
and 0.23247(43) respectively; they differ by 2.9$\sigma$. The corresponding numbers for 1998
 are 0.23101(31) and 0.23223(38) respectively, with a difference of 2.5$\sigma$. The shifts
 are +0.9$\sigma$ and -0.6$\sigma$ respectively. The largest change from a single
 experiment between 1996 and 1998 occurs in the ALEPH \tpol~value of $\sleff$
 which changes
 from 0.2333(14) to 0.23146(58), a -1.3$\sigma$ shift. The overall consistency C.L. of the
 leptonic data shown in Fig.1. improves from 8.2$\%$ to 32$\%$ between 1996 and 1998}
shown in Fig.1.
 In fact the values of $\sleff$ derived from $A_{LR}$, $A_{FB}^{0,e}$, $A_{FB}^{0,\mu}$, 
 $A_{FB}^{0,\tau}$ and the L3 \tpol~measurement of $A_l$ were all consistent.
 On the other hand, systematically higher values were given by the \tpol~ measurements
 of ALEPH, DELPHI and OPAL and by the precise  $A_{FB}^{0,b}$ measurement, on the
 assumption of SM values for the b quark couplings. This consistency of the
 majority of the determinations of $\sleff$ (including $A_{LR}$) was obscured by the
 ordering of the data in the cumulative averages of $\sleff$ shown in Fig.5
 of Ref.[1]. If $A_{FB}^{0,b}$ had been the last datum to be included instead of
 $A_{LR}$, the former not the latter would apparently have given the largest single
 contribution to the $\chi^2$. The model independent analysis of Refs.[4,5]
 showed that, indeed, it is in the b quark couplings that the largest apparent deviation
 from the SM predictions are found. Repeating the analysis described above for the 
 data set of Ref.[15] leads to 95$\%$ C.L. upper limits on $m_H$ of $101_{-43~-25}^{+77~+48}$
 GeV or $220_{~-92~-52}^{+114~+71}$ GeV when the b quark 
 data are, respectively, excluded or included. These limits may be compared with the C.L.,
 based on the maximum of the 1$\sigma$ contour in the $m_t$ versus $m_H$ plane ( after 
 rescaling the errors according to the PDG recipe of Ref.[16]), of
 650 GeV, quoted in Ref.[1]. The limit similarly defined from the 1$\sigma$ contour
 of the fit to the 1996 data, similar to that presented in the fifth row of Table 6, is
 much lower, 249 GeV.  
 Also the `maximal' 95$\%$ C.L., calculated as described above, but using all the 
 corresponding 1996 data
 is only 405 GeV. 
\par Neither Ref.[1] nor the more recent papers of Chanowitz~\cite{x3} discuss either the 
 sensitivity of the Higgs mass limits to the b decay data or {\it the goodness of fit
 of the SM to the subsets of data directly sensitive to $m_t$ and $m_H$}.
 The $\chi^2$ quoted for the fit to all data of Ref.[1] was 19 for 14 $dof$ (C.L.=0.17).
 However, of this $\chi^2$, 9.9 (i.e. 52$\%$) is contributed by only three data:
 $R_b$, $A_{FB}^{0,b}$ from LEP and $A_b$ from SLD, just those that are directly
 sensitive to the b quark couplings. \footnote{ The situation is similar for the
 fit to the most recent data~\cite{x6}. The same three data contribute 8.7
 out of a total $\chi^2$ of 16.4} The C.L. of a $\chi^2$ of 9.9 for 3$dof$ is only
 0.02. In fact, the fits of Ref.[1] include many data that are quite insensitive 
 to $m_t$ and $m_H$. 
 The poor quality of the SM fit to the extracted effective couplings due
 to the apparently anomalous values for the b quarks
 \footnote{The large deviations of the b quark couplings from the SM predictions
  have also been reported, but not discussed in detail, by
 Renton~\cite{x17} and Grant and Takeuchi~\cite{x18}}  was first pointed out in Refs.[4,5].
\par There are three possible causes of the apparent deviation of the b quark couplings 
from the SM predictions: (i) a statistical fluctuation, (ii) a hitherto unknown systematic 
error in the LEP $\Afbb$ and SLD $A_b$ measurements, (iii) new physics, beyond the SM,
in b decays. For example, the deviations of the $\sin^2\theta_{eff}^{lept}$ values 
derived from $A_{FB}^{0,b}$ and some tau-polarisation measurements from those derived
from purely leptonic observables (see Fig 1) has been conjectured to arise from 
poorly understood hadronisation effects~\cite{x19}. A `new physics' scenario has been
proposed~\cite{x20} that explains the anomalous b quark couplings while leaving 
unchanged the Standard Model $m_H$ dependence of the leptonic effective couplings. 
\par It can be seen from Table 3 that all the apparent deviation is in the single
 parameter $A_b$. The parameter $\sbb$ (derived from $R_b$) differs from the
SM expectation by only 1.1$\sigma$. This good agreement of $\sbb$ is highly
 non-trivial in the SM due to the large flavour specific and $m_t$ dependent vertex
 corrections in Z decays to b quarks.
Removing these corrections results in a predicted value of $\sbb$ of 0.370, which 
differs by 2.7$\sigma$ from the measured value quoted in Table 3. It seems unlikely
that the good agreement observed for $\sbb$ with the SM prediction is a lucky accident,
 as must
 be the case if there is indeed a  new physics explanation of the measured values of
 $\abb$ and $\vbb$ shown in Table 4. This argument disfavours the explanation (iii)
 and favours (i) or (ii) (or some combination of the two).
 \par Assuming gaussian errors,
 the probability that one of the six parameters in Table 3 shows a deviation from the
 SM prediction of
 3$\sigma$, or greater, is 1.6$\%$. In order to obtain a value of  $\sleff$ 
 equal to the weighted average of the leptonic measurements shown in Fig. 1, the value
 of $\Afbb$ must be increased by 5.5$\%$, as compared to the estimated fractional 
systematic error on $\Afbb$ of
1.0$\%$. There is a correction of $\simeq$1.9$\%$ to the raw measured value of $\Afbb$ to account
for QCD effects~\cite{x21}. To explain the observed deviation of $\sleff$
as a QCD effect, an extra 
correction of 3.6$\%$,  would be needed 
 as compared to the estimated systematic error on the QCD correction of only 0.3$\%$.
\par Independently of the explanations (i),(ii),(iii) of the apparently anomalous
  b quark couplings, the upper limit on the Higgs boson mass is expected, in all
cases,be lower than those hitherto quoted~\cite{x1,x6,x14,x15,x22} derived from fits
 including the 
 b quark data. In cases (i) and (ii), the true value of $\sleff$ should be consistent
with the average of the leptonic data shown in Fig 1, leading to a limit similar
to those found in the lepton and lepton + c quark fits shown in Tables 5 and 6. 
 In case (iii), where the SM breaks down for the b quark couplings, the latter can 
give no information whatever \footnote{As pointed out above, also in the
case that the SM is respected, the b quark effective couplings have only a very weak
 sensiivity to $m_H$.} on the Higgs mass,
 so, assuming the SM does apply to the
 lepton and c quark
data (as is the case for the model described in Reference[20]), the same conclusion
 is drawn about the Higgs mass limit. Actually, the largest
uncertainty on the limit is currently due the the errors of the $\alpha(M_Z)$
 determination. Using instead of the value quoted in Table 2, derived from Ref.[23],
 the more recent, but model dependent, determination of Ref.[24]
 of $\alpha(M_Z)^{-1} = 128.923(36)$ leads to the
 tighter 95$\%$ C.L. upper bounds on $m_H$ of $121_{-23~-29}^{+31~+42}$ GeV
 ($176_{-30~-38}^{+37~+49}$ GeV)
 in the case that the 
b quark data is excluded (included).
\par In conclusion, it may be stated that in view of the still large uncertainties 
on the upper bound on $m_H$ due to the experimental errors on $\alpha(M_Z)$ 
and $m_t$ there is no conflict with the current 95$\%$ C.L. lower limit of 89.8 GeV on
the $m_H$~\cite{x13}. However, independently of the interpretation given to the
apparent anomalies in the b quark couplings, the most recent data indicates 
that the mass of the Higgs boson, if it exists, is probably less than 200 GeV.
\par After completion of the first version of this paper several other global
 analyses of the same data set have been performed~\cite{x25,x26,x27}. Erler and
 Langacker~\cite{x25} did discuss the 3$\sigma$ deviation in the LEP+SLD average
 value of $A_b$. However, the effective couplings of the heavy quarks were not
 extracted,
 and, on the basis of the good $\chi^2$ of a global fit to a total of 42 data, they
 concluded that `the fit to all precision data is perfect'. They did not notice,
 that as pointed out in this paper, a fit to the effective couplings alone ( although
 the `information content' concerning $m_t$ and $m_H$ is essentially equivalent to the
 data used in their fit), gives only poor agreement with the SM. The effect of the
 apparent $A_b$
 anomaly on the $\chi^2$ of their global fit is masked by the contributions of many data
 that have large errors and/or are relatively insensitive to $m_t$ and $m_H$. 
 Neubert~\cite{x25}
 mentioned the $A_b$ deviation, and although discussing the sensitivity of $m_H$
 to different measurements, did not point out in the text the strong 
 sensitivity to $A_{FB}^{0,b}$.
  As shown in the present paper, this increases the fitted value of $m_H$, even
 though, in the SM, the effective b quark couplings are quite insensitive to $m_H$.
 In fact, inspection of Fig.5. of~\cite{x25} shows clearly that the determination of
 $m_H$ from $A_{FB}^{0,b}$ alone is strongly biased towards high values. This however
 was not remarked, and the     
 author concluded, instead, that `there is no particular set of measurements that pulls
 $m_H$ down'. Renton~\cite{x27} specifically discussed the apparent anomaly in the
 b quark data and extracted the effective couplings, finding results in good agreement
 with those quoted in Table 4 above. The sensitivity of the Higgs mass to inclusion or 
 exclusion
 of the the b quark data was also studied and results consistent with those shown in
 the present paper obtained.
\par 
{\bf Acknowledgements}
\par  I thank M.Consoli, M.Dittmar and Tariq Aziz for discussions, and J.Mnich for help with the
ZFITTER program package.
\pagebreak

\pagebreak
\begin{table}
\begin{center}
\begin{tabular}{|c|c|c|c|c|c|c|} \hline
\multicolumn{7}{|c|}{LEP} \\ \hline
\multicolumn{3}{|c}{leptons} & \multicolumn{2}{|c}{ c quarks} &
\multicolumn{2}{|c|}{ b quarks} \\ \hline
 $\Afbl$ & $A_l$ ($\tau$ poln.) & $\Gamma_l$ (MeV) & $\Afbc$ & $R_c$ &
$\Afbb$ & $R_b$   \\  \hline
 0.01683(96) & 0.1452(34) & 83.90(10) & 0.0714(44) & 0.1733(44) &
 0.0991(21) & 0.21656(74)  \\ \hline
\multicolumn{6}{|c|}{SLD} \\ \cline{1-6}
\multicolumn{2}{|c}{leptons} & \multicolumn{2}{|c}{c quarks} &
  \multicolumn{2}{|c}{b quarks} &  \multicolumn{1}{|c}{  } \\ \cline{1-6}
\multicolumn{2}{|c}{ $A_{LR}$} & \multicolumn{2}{|c}{ $A_c$ } &
  \multicolumn{2}{|c}{ $A_b$ } &  \multicolumn{1}{|c}{  } \\ \cline{1-6}
\multicolumn{2}{|c}{ 0.1511(24) } & \multicolumn{2}{|c}{ 0.638(40) } &
  \multicolumn{2}{|c}{ 0.856(36) } & \multicolumn{1}{|c}{  } \\ \cline{1-6}
\cline{1-6}
\end{tabular}  
\caption[]{ Average values of electroweak observables~\cite{x6} directly sensitive
 to $m_H$ and $m_t$  used
 in the analysis. The SLD measurements of $R_c$ and $R_b$ are included in the
`LEP' averages shown }
\end{center}
\end{table} 
\begin{table}
\begin{center}
\begin{tabular}{|c|c|c|c|c|c|} \hline
  $M_Z$ GeV  & $\Gamma_Z$ (GeV)  & $\sigma_h^0$ (nb)  & $R_l$ & $\alpha(M_Z)^{-1}$ &
 $G_{\mu}$ (GeV$^{-2}$)  \\  
\hline
91.1867(21) & 2.4939(24) & 41.491(58) & 20.765(26) & 128.896(90) & 1.16639(1)$\times 10^{-5}$ \\
\hline
\end{tabular}
\caption[]{ Other electroweak parameters used in the analysis~\cite{x6,x12}. }  
\end{center}
\end{table}
\begin{table}
\begin{center}
\begin{tabular}{|c|c|c|c|c|c|c|} \cline{2-7}
\multicolumn{1}{c}{ } & \multicolumn{2}{|c}{leptons } & \multicolumn{2}{|c}{ c quarks }
 & \multicolumn{2}{|c|}{ b quarks }  \\ \cline{2-7}
\multicolumn{1}{c|}{ } & $A_l$  & $\sbl$  &  $A_c$ & $\sbc$  & $A_b$ & $\sbb$  \\ \hline
 Meas. & 0.1492(18) & 0.25243(30)  & 0.638(28) & 0.2950(74) &
0.878(19)  & 0.3662(14) \\ \hline
 SM &  0.1467 & 0.25272  & 0.6677 & 0.2882 & 0.9347 & 0.3647 \\ \hline
Dev.($\sigma$) & 1.4 & -1.0  &-1.1  &  0.92 & -3.0 & 1.1 \\
 \hline      
\end{tabular}
\caption[]{ Measured values of $A_f$ and $\sbf$ compared to SM 
 predictions for $m_t =$ 174 GeV, $m_H =$ 100 GeV.  Dev($\sigma$) = (Meas.-SM)/Error. } 
\end{center}
\end{table}
\begin{table}
\begin{center}
\begin{tabular}{|c|c|c|c|c|c|c|} \hline
\multicolumn{1}{|c}{ } & \multicolumn{2}{|c}{leptons } & \multicolumn{2}{|c}{ c quarks }
 & \multicolumn{2}{|c|}{ b quarks }  \\ \cline{2-7}
      & $\abl$  & $\vbl$  &  $\abc$ & $\vbc$ & $\abb$ & $\vbb$  \\ \hline
 Meas. & -0.50102(30) & -0.03759(46) & 0.511(7) & 0.184(10) &
 -0.5208(66) & -0.311(11) \\ \hline
 SM & -0.50136 & -0.03697 & 0.501 & 0.191 & -0.4981 & -0.3434 \\ \hline
Dev.($\sigma$) & 1.1 & -1.3 & 1.4 & -0.7 & -3.4 & 2.9 \\
 \hline     
\end{tabular}
\caption[]{ Measured values of effective coupling constants
 compared to SM 
 predictions for $m_t =$ 174 GeV, $m_H =$ 100 GeV.  Dev($\sigma$) = (Meas.-SM)/Error. } 
\end{center}
\end{table}
\begin{table}
\begin{center}
\begin{tabular}{|c|c|c|c|c|c|} \hline
\multicolumn{1}{|c}{ } & \multicolumn{3}{|c}{ $m_H$ free  } &
 \multicolumn{2}{|c|}{ $m_H = 200$ GeV  }  \\ \hline
Fitted Quantities & $m_t$ (GeV) & $m_H$ (GeV) & C.L.($\%$) &
 $m_t$ (GeV) & C.L.($\%$) \\ \hline
 $A_l$, $\sbl$ &$164.6_{-6.2}^{+5.8}$ & $25.8_{-10.9}^{+20.5}$ [63] &  100 &
  $184.8_{-5.5}^{+5.4}$ &  2.5 \\ \hline
 $A_l$, $\sbl$, $A_c$, $\sbc$ & $164.4_{-6.1}^{+5.9}$ & $28.2_{-12.0}^{+20.8}$ [66] & 56
 & $184.5_{-5.5}^{+5.3}$ &  16 \\ \hline
 $A_l$, $\sbl$, $A_b$, $\sbb$ & $161.5_{-5.7}^{+5.5}$ & $41.9_{-16.6}^{+19.8}$ [79] & 1.3
 &  $180.8_{-5.1}^{+5.0}$ &  0.3 \\ \hline
 $A_l$, $\sbl$, $A_c$, $\sbc$  $A_b$, $\sbb$ & $161.6_{-5.6}^{+5.4}$ &
 $45.1_{-16.8}^{+20.5}$ [83] & 3.9
 & $180.4_{-5.0}^{+4.9}$ & 2.0 \\ \hline 
\end{tabular}
\caption[]{ SM fits to different data sets. 95$\%$ C.L. upper limits for $m_H$ are 
given in square brackets.} 
\end{center}
\end{table}
\begin{table}
\begin{center}
\begin{tabular}{|c|c|c|c|c|} \hline
Fitted Quantities & $\alpha(M_Z)^{-1}$  & $m_t$ (GeV) & $m_H$ (GeV) & C.L.($\%$)  \\ \hline
\multicolumn{1}{|c|}{ } & 128.986 & $171.7_{-3.8}^{+3.7}$ &
 $81.6_{-31.6}^{+48.8}$ [175] & 51  \\ \cline{2-5}
$A_l$, $\sbl$, $A_c$, $\sbc$, $m_t$ & 128.896  & $171.2_{-3.8}^{+3.7}$ &
$47.2_{-24.5}^{+29.8}$ [105] & 53  \\ \cline{2-5}
\multicolumn{1}{|c|}{ } & 128.806 &  $171.8_{-3.8}^{+3.7}$ &
 $21.7_{-9.1}^{+20.1}$ [59] & 65 \\ \hline
\multicolumn{1}{|c|}{ } & 128.986 & $171.9\pm 3.6$ &
 $131.3_{-41.6}^{+58.9}$ [240] & 4.5  \\ \cline{2-5}
 $A_l$, $\sbl$, $A_c$, $\sbc$, $A_b$, $\sbb$, $m_t$ & 128.896 &
 $171.5 \pm 3.6$ & $83.3_{-27.1}^{+39.1}$ [156] & 4.3 \\ \cline{2-5}
\multicolumn{1}{|c|}{ } & 128.806 &  
 $171.4 \pm 3.6$ & $47.9_{-19.3}^{+23.2}$ [92] & 4.4 \\ \hline 
\end{tabular}
\caption[]{ SM fits to different data sets. 95$\%$ C.L. upper limits for $m_H$ are 
given in square brackets. } 
\end{center}
\end{table}
\newpage
\begin{figure}[htbp]
\begin{center}\hspace*{-0.5cm}\mbox{
\epsfysize10.0cm\epsffile{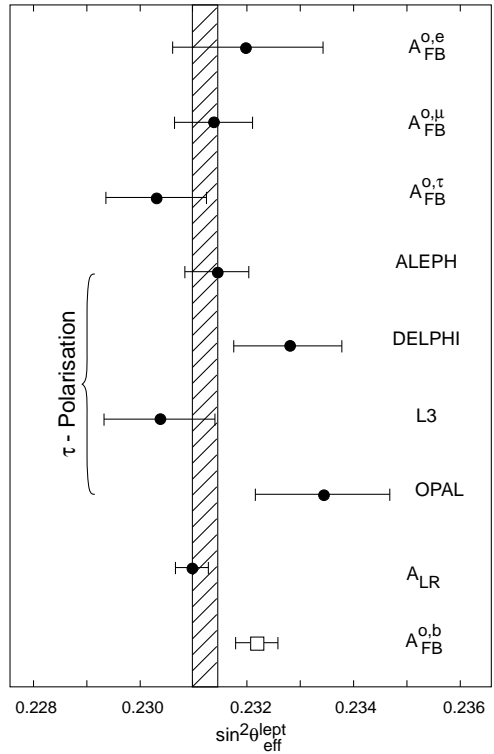}}
\caption{ Values of $\sleff$ calculated using different electroweak observables.
 The cross-hatched band shows the $\pm 1 \sigma$ region of the weighted average
 of the leptonic measurements (solid circles). The value derived from $\Afbb$
 (open square) assuming SM values for the b quark couplings and not included
 in the weighted average, is shown for comparison.}
\label{fig-fig1}
\end{center}
 \end{figure}  
\begin{figure}[htbp]
\begin{center}\hspace*{-0.5cm}\mbox{
\epsfysize10.0cm\epsffile{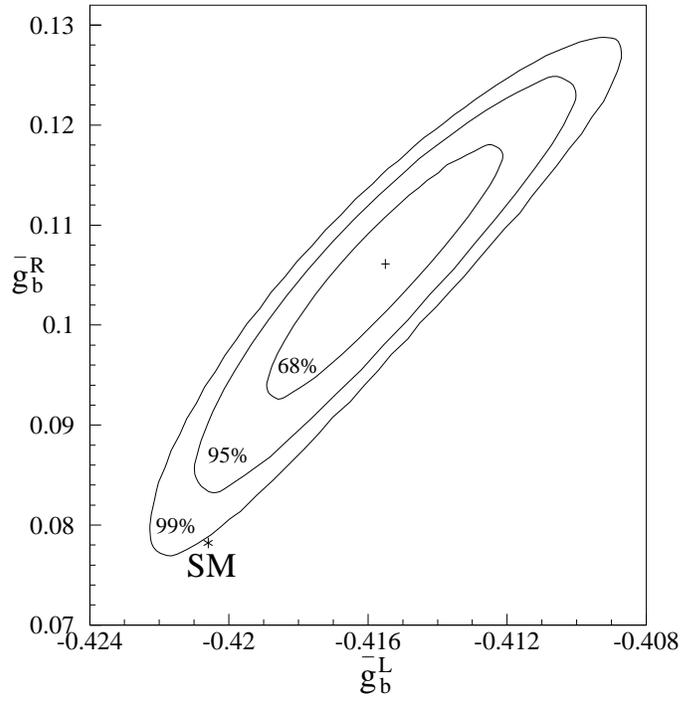}}
\caption{ Contour plot for fit to $\gbl$ and $\gbr$. The 68$\%$, 95$\%$ and
 99$\%$ C.L. contours are shown. The SM prediction is for $m_t = 174$ GeV,
 $m_H = 100$ GeV.}
\label{fig-fig2}
\end{center}
 \end{figure}  
\begin{figure}[htbp]
\begin{center}\hspace*{-0.5cm}\mbox{
\epsfysize10.0cm\epsffile{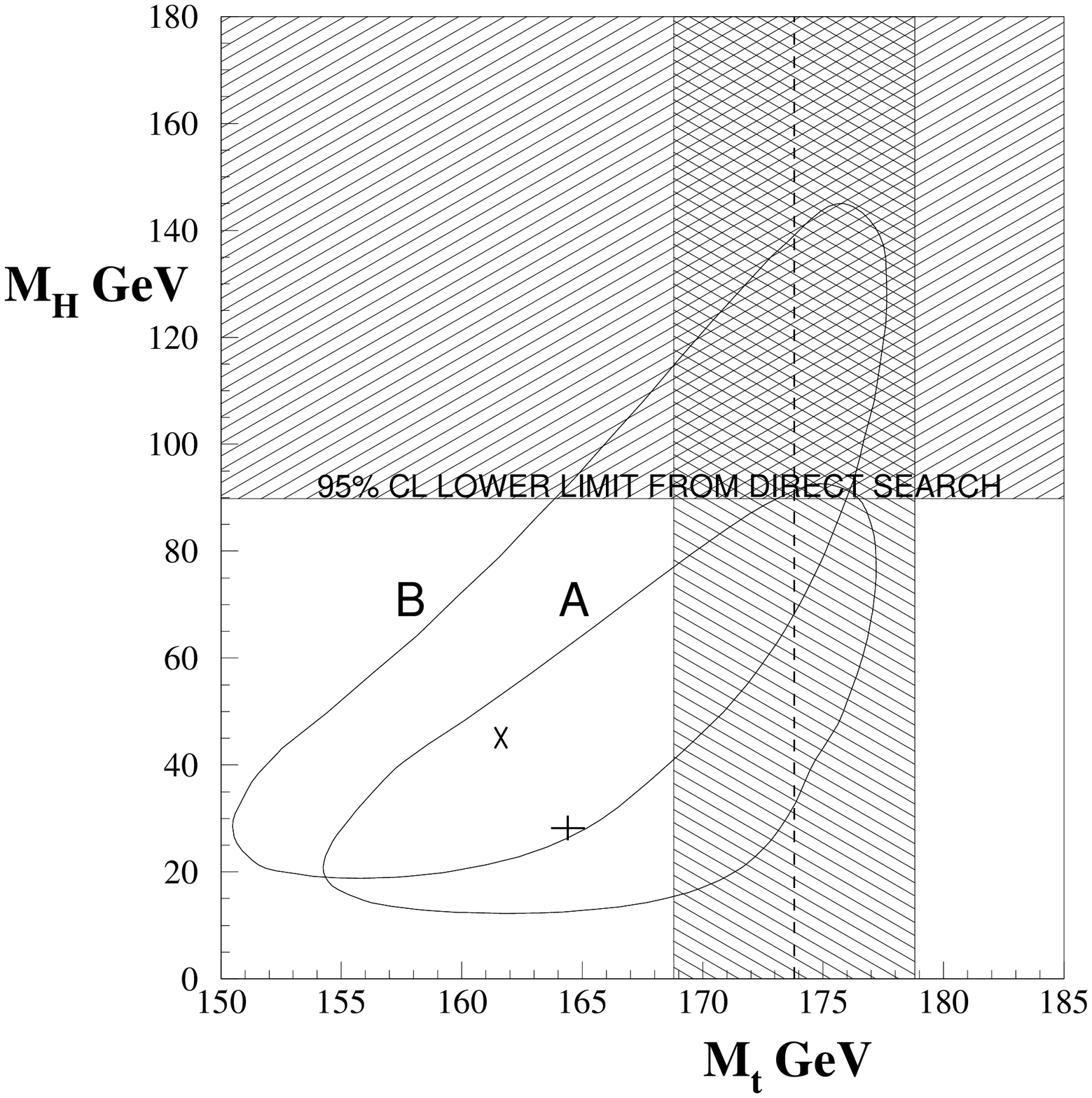}}
\caption{ Contour plot for fit to $m_t$ and $m_H$. Contours at the 68$\%$ C.L. 
 are shown for fits to the lepton and c quark data only (Contour A, best fit position 
 $+$) and to lepton, c quark and b quark data (Contour B, best fit position $\times$).
 The cross hatched areas show the allowed regions for the direct Higgs search~\cite{x13}
 and the direct measurement: $m_t=173.8 \pm 5.0$ GeV~\cite{x6}. These limits are shown for
 comparison only, and are not included in the fits.}
\label{fig-fig3}
\end{center}
 \end{figure} 

\begin{thebibliography}{99}
\bibitem{x1}
A.Gurtu, Phys. Lett. {\bf B368}, 247 (1996).
\bibitem{x2}
S.Dittmaier, D.Schildknecht and G.Weiglen, {\bf B385}, 415 (1996).
\bibitem{x3}
M.S.Chanowitz, Phys. Rev. Letters {\bf 80}, 2521 (1998),
and pre-print LBNL-42103 (hep-ph/9807452).
\bibitem{x4}
J.H.Field, Mod. Phys. Lett. A, Vol. 13, No. 24, 1937  (1998).
\bibitem{x5}
J.H.Field, Phys. Rev. {\bf D58} 093010-1 (1998).
\bibitem{x6}
M.Gr\"{u}newald, in proceedings of the XXIX International
 Conference on High Energy Physics, UBC, Vancouver, BC, Canada, 
July 23-29 1998.  
\bibitem{x7}
G.Degrassi, P.Gambino  and A.Vicini, Phys. Lett. {\bf B383}, 219 (1996),
G.Degrassi, P.Gambino  and A.Sirlin, Phys. Lett. {\bf B394}, 188 (1997),
G.Degrassi, et al. Phys. Lett. {\bf B418}, 209 (1998),
\bibitem{x8}
D.Bardin et al. FORTRAN package ZFITTER, Preprint CERN-TH 6443/92.
\bibitem{x9}
G.Rodrigo, Nucl. Phys. B (Proc. Suppl.) {\bf 54A}, 60 (1997).
\bibitem{x10}
M.Schmelling in Proceedings of the 28th International
Conference on High Energy Physics, 
Warsaw 1996, Eds. Z.Ajduk and A.K.Wroblewski.
\bibitem{x11}
P.N.Burrows, Talk presented at the 3rd International Symposium 
on Radiative Corrections, August 1-5 1996, Cracow, Poland.
Pre-print: SLAC-PUB-7293.
\bibitem{x12}
Review of Particle Properties, Particle Data Group,
C.Caso et al. Eur. Phys. J.  {\bf C3} 1 (1998).
\bibitem{x13}
ALEPH, DELPHI, L3 and OPAL Collaborations, 
`Lower bound on the Standard Model Higgs boson mass 
 from combining the results of the four LEP experiments',
 CERN-EP/98-046.
\bibitem{x14}
The LEP Collaborations ALEPH, DELPHI, L3, OPAL,
the LEP Electroweak Working Group and the SLD
Heavy Flavour Group. CERN PPE/95-172 (1995). 
\bibitem{x15}
The LEP-SLD Electroweak Working Group (see Ref.[14]),
 CERN-PPE/96-183 (1996).
\bibitem{x16}
Review of Particle Properties, Particle Data Group
L.Montanet et al. Phys. Rev. {\bf D 50}, 1173 (1994).
\bibitem{x17}
P.B.Renton, Int. Journ. Mod. Phys. {\bf A12}, 4109 (1997).
\bibitem{x18}
\mbox{A.K.Grant and T.Takeuchi,}\mbox{`An analysis of Precision Electroweak}     
 Measurements:\mbox{ Summer 1998 Update', VPI-IPPAP-98-1}, UCLA/98/TEP/20,
 hep-ph/9807413.
\bibitem{x19}
T.Aziz, Mod. Phys. Lett. {\bf A12}, 2535 (1997).
\bibitem{x20}
D.Chang, W.-F.Chang and E.Ma {\it Alternative Interpretation of the Tevatron Top
 Events} UCRHEP-T237, hep-ph/9810531.
\bibitem{x21}
D.Abbaneo et al. `QCD Corrections to the forward-backward asymmeties
of c and b quarks at the Z pole'. CERN-EP/98-32.  
\bibitem{x22}
The\mbox{ LEP-SLD Electroweak Working Group}
\mbox{(see Ref.[14]),}\mbox{ LEPEWWG/97-01 (1997).}
\bibitem{x23}
S.Eidelmann and F.Jegerlehner, Z. Phys. {\bf C67}, 585 (1995).
\bibitem{x24}
M.Davier and A.H\"{o}cker, Phys. Lett. {\bf B419}, 419 (1998).
\bibitem{x25}
J.Erler and P.Langacker, {\it `Status of the Standard Model'}, Pre-print UPR-0816-T,
hep-ph/9809352.
\bibitem{x26}
F.Teubert, {\it Precision test of the Standard Model from Z physics},
hep-ph/9811414.
\bibitem{x27}
P.B.Renton, {\it Are there anomalous Z fermion couplings?} Pre-print OUNP-98-08,
hep-ph/9811415. 
\end{thebibliography}
\end{document}